# Nuclear spin induced oscillatory current in spin-blocked quantum dots

Keiji Ono[1] and Seigo Tarucha[1, 2, 3]

[1]*Dept. of Physics, University of Tokyo, 7-3-1 Hongo Bunkyo-ku, Tokyo 113-0033, Japan*
*Phone: +81-3-5841-4162 Fax: +81-3-5841-4162 E-mail: ono@phys.s.u-tokyo.ac.jp*
[2]*NTT Basic Research Laboratories, Wakamiya 3-1 Morinosato Atsugi-shi, Kanagawa 243-0198, Japan*
[3]*ERATO Mesoscopic Correlation Project, Japan Science and Technology Corporation,*
*Wakamiya 3-1 Morinosato Atsugi-shi, Kanagawa 243-0198, Japan*

Hyperfine coupling of electron spins to nuclear spins is studied for a GaAs-based double quantum dot in the spin blockade regime where the electron conduction is mostly blocked by Pauli effect unless the electron spin state in the double dot is changed. A current flowing through the double dot shows time-dependent oscillations with a period of as long as 200 sec in a certain DC magnetic field range. The oscillatory behavior is significantly diminished by application of an AC magnetic field whose frequency can induce nuclear magnetic resonance for $^{71}$Ga and $^{69}$Ga, respectively. A possible nuclear spin polarization mechanism due to hyperfine flip-flop scattering is proposed.

Electronic properties of semiconductor quantum dots are often strongly influenced by spin-related interactions such as Hund's coupling, Pauli exclusion and the Kondo effect [1-3]. These spin effects are associated with not only the ground state but also the excited states, and revealed by measurements of electron transport. This implies that the spin configuration is robust on the time scale much longer than the transport time through the quantum dot system, that is, the degree of electron spin freedom is well isolated from the environment [4]. Consistently theory predicts that spin-orbit coupling accompanied by phonon scattering and hyperfine coupling to nuclei are the only possible but very weak spin scattering sources in GaAs-based quantum dots [5,6].

Electron spin and nuclear spin degrees of freedom in low-dimensional solid-state system are subject to intensive studies from the viewpoints of applications to future spintronics and quantum computations [7]. Hyperfine interactions of electron spins to nuclear spins can play an important role in these applications. However, no experiments on the hyperfine interactions are reported for quantum dots to date but for a two-dimensional electron gas in the quantum Hall regime [8]. In addition, studies on quantum dot systems can give us new insight to the hyperfine interactions, because the related 0D electronic configuration is well defined. In this Letter we present experimental studies on hyperfine interactions for a two-electron spin triplet state confined in a double quantum dot system. Formation of such a triplet state blocks a single electron tunneling current flow through the double dot system by Pauli exclusion, and an excess current is observed when the spin triplet state suffers from spin-flip transitions. Contributions from nuclei to the excess current are explored using a technique like nuclear magnetic resonance (NMR) but detected by a single electron tunneling current with much higher sensitivity. This is a first demonstration of an electrically detected NMR of quantum dots. We propose a model that at a certain magnetic field the nuclear spins can be dynamically polarized due to the hyperfine flip-flop interaction.

One of the key ingredients in this work is to prepare an excited but long-lived spin triplet state in a double dot system. We have recently observed a current rectification effect due to Pauli exclusion using a device such that two quantum dots (dot 1 and 2) are weakly connected in series between a source and drain contacts (see Fig.1) [9]. When the electrostatic potentials for two dots are tuned such that the three charge states of $(N_1, N_2)$ = (0,1), (1,1) and (0,2) are degenerated at zero source drain voltage, $V_S$ = 0 V, just lifting the Coulomb blockade. Here $N_1$ ($N_2$) is the number of electrons in dot 1 (dot 2). For (1,1), two spin-states, i.e. a spin singlet and triplet states are present but only slightly energetically spaced. On the other hand, the (0,2) state only takes a spin singlet state because two electrons share the same lowest $1s$ orbital state of dot 2 due to Pauli exclusion. The second lowest $2p$ orbital state is located well above (~ 5 meV) the lowest $1s$ state in our quantum dot, so the occupation of a (0,2) triplet state (i.e. $1s2p$ filling) can be neglected for a small $V_S$ (< 10 mV) at low temperatures (< 2 K) [9]. By applying a small but finite negative $V_S$ ("forward bias" in the current rectification), electron current is carried by cycles of three irreversible (inelastic) transitions of $(0,1) \rightarrow (0,2)$ singlet $\rightarrow (1,1)$ singlet $\rightarrow (0,1)$ … For a positive $V_S$ ("reveres bias"), however, although there are electron current carrying cycles of $(0,1) \rightarrow (1,1)$ singlet $\rightarrow (0,2)$ singlet $\rightarrow (0,1)$ …, once another transition of $(0,1) \rightarrow (1,1)$ triplet takes place, further electron transfer is prohibited by combined Coulomb blockade and Pauli exclusion before the spin-flip relaxation of the triplet occurs ("spin blockade") .

To set up the spin blockade (SB) condition we use a vertical double dot device, which is essentially the same as used for our previous work: a gated sub-micron pillar of a triple barrier resonant tunneling structure composed of two 8 nm thick $Al_{0.22}Ga_{0.78}As$ outer barriers, a 6 nm thick $Al_{0.22}Ga_{0.78}As$ center barrier and two 12 nm thick $In_{0.05}Ga_{0.95}As$ wells [9,10]. For comparison a similar device but having a 7.5 nm thick center barrier is prepared.

Figure 1 shows the current, $I$, flowing through the double dot versus voltage, $V_S$, measured at 1.8 K. The gate voltage, $V_G$, is fixed at the position of the Coulomb peak at $V_G = 0.05$ V in the linear conductance (Lower right inset). Then electrons are transported through the two-electron states in the double dot [4]. The SB region appears in the $V_S$ range from 2 to 6 mV, where a small leakage current of $I \sim 1$ pA is observed. This means the $(1,1)$ triplet has a finite lifetime of $e/I \sim 100$ ns ($e$, elementary charge), which is much longer than the inelastic electron tunneling time ($\sim$ a few ns) throughout the system. The leakage current can arise from spin scattering events that change the $(1,1)$ triplet state to the $(1,1)$ singlet state and co-tunneling processes that lead the $(1,1)$ triplet directly to the $(0,1)$ state [9].

We set $V_S$ at 3.0 mV in the SB region of Fig. 1, and measure the leakage current as a function of DC magnetic field, $B_{DC}$, applied horizontally. Figure 2 (a) shows the data measured for $B_{DC}$ from 0 T to 1.2 T with a constant sweep rate of 5 hours/T. The applied $B_{DC}$ field is sufficiently low that neither of the changes in the orbital energy and in the effective thickness of barriers can be neglected. As the magnetic field initially increases, the current is nearly constant (< 2 pA) for $B_{DC}$ < 0.5 T, and rises with a sharp step at $B_{DC} \sim 0.5$ T. Then it starts to fluctuate more strongly with increasing $B_{DC}$ up to $\sim$ 0.87 T, and suddenly decreases for $B_{DC}$ > 0.9 T. A similar characteristic, i.e. a step followed by fluctuations, are observed at different values of $V_S$ and $V_G$ within the SB region. Note we observe a shift of the $I$-$B_{DC}$ curve to the lower field by about 0.2 T when the $B_{DC}$ is swept down after it is swept up beyond 1 T. For any $B_{DC}$ field in the current fluctuation regime (0.6 $\sim$ 0.87 T), the current shows periodic oscillations as a function of time (Fig. 2(b)). Both of the period and amplitude of the current oscillations increase with $B_{DC}$, and become maximal with the period of as long as $\sim$ 200 sec and the amplitude of $\sim$ 0.4 pA near $B_{DC} = 0.87$ T. These oscillations last with no definite damping for 15 hours or longer. We observe no clear periodic oscillations after the current decreases to the low level for $B_{DC}$ > 0.87 T, although there still remain significant fluctuations of order 0.1pA, which are much larger than the noise level of $\sim$ 10 fA. Note that in Fig. 2(b) variations of the current slower than our measurement time constant of $\sim$ 1 sec can only be detected.

$I$ - $B_{DC}$ characteristics similar to that shown in Fig. 2(a) are observed for four double dot samples: three samples with a 6 nm center barrier and one sample with a 7.5 nm center barrier. A current step and oscillations are observed in the smaller $B_{DC}$ field range (a step at $\sim$ 0.3 T and maximal oscillations at $\sim$ 0.6 T) in the device with a 7.5-nm center barrier. Other two 6-nm-center-barrier samples show a similar step and hysteresis, although the oscillations are less clear.

To investigate how the current oscillations initially evolve with time in the SB region, we have performed the following transient measurements. First we set $B_{DC} = 0.87$ T and $V_S = 3.0$ mV, where the current oscillations have a nearly maximal period ($\sim$ 200 sec) and amplitude ($\sim$ 0.4 pA). Regions of Coulomb blockade with $N_1 + N_2 = 2$, and SB are present for $|V_S| < 1$ mV and $V_S > 1$ mV, respectively [9]. Then, we switch $V_S$ from 3.0 mV to $-1$ mV, where the Coulomb blockade is almost lifted and a small current of $\sim -1$ pA flows. Thus after dwelling for 10 min outside the SB region we switch back $V_S$ to 3.0 mV. Then, for the first 5 min we observe current oscillations with a short period and small amplitude but both gradually increasing with time until approaching the steady state oscillatory condition. The transient behavior before approaching the steady state condition becomes less clear for the shorter dwell time. This measurement clearly shows that the time scale of $\sim$ 5 min is needed for both establishing and erasing the oscillatory behavior. Such a slow response can be associated with nuclear spin system, which has an unusually long longitudinal decay time constant longer than 10 min at low temperatures [8].

As is well known for a NMR technique, nuclear spin effects can be confirmed from response to an AC magnetic field. We use a three-turn coil of 3-mm diameter located 0.5 mm above the device to apply vertically an AC magnetic field, $B_{AC}$, to the double dot (see Fig. 1), and measure the change in the oscillatory current. Figure 3(a) shows the data measured for various frequencies of $B_{AC}$ at $B_{DC} = 0.85$ T. A strong reduction in both of the oscillation period and amplitude of the current is observed when the frequency of the $B_{AC}$ field matches to [71]Ga nuclear spin resonance (Fig. 3(b)). The resonance frequency changes linearly with $B_{DC}$ (Inset to Fig. 3(b)). A similar behavior is observed for $^{69}$Ga resonance, although the signal is smaller, i.e. the change in the oscillation period is smaller. So far we have not observed any signal associated with the As, In and Al nuclei.

Both of the current step and oscillations are observed only within the SB region. Our transport measurement is insensitive to the change in the nuclear spin state of the bulk GaAs regions (contact leads). Nuclear spin polarization only due to the nuclear Zeeman energy is so small in our measurement condition ($10^{-4}$ at 1.8K and 0.85T) that it hardly influences electron transport. Therefore, we consider that the nuclei in *quantum dots*, of which number is of the order of $10^5$, are dynamically polarized at a certain $B_{DC}$ field in the SB region. Note the Overhauser effect may affect the electron spin and the transport characteristics as well [8].

Here we propose a tentative model that accounts for the dynamic polarization of nuclei in the SB double dot at a certain $B_{DC}$ field. Transition from the (1,1) triplet to (1,1) singlet can be induced by hyperfine flip-flop scattering with the nuclei in the quantum dots. However, it is pointed out that the flip-flop is significantly suppressed due to the discreteness of electron energy in quantum dots [6]. A small but finite tunnel coupling and exchange interaction between dots lift the degeneracy of the (1,1) singlet and triplet states. The energy separation between these states is calculated to be 10-40 μeV for the 6-nm-center-barrier sample [11]. This energy separation decreases with increasing $B_{DC}$ field, and finally one of the Zeeman-split (1,1) triplet states having $S_Z = +1$ and the (1,1) singlet state become degenerate. Then the hyperfine flip-flop scattering only turns on between these two states and not for the other triplet states having $S_Z = 0$ and $-1$. Thus an electron spin-flip always provides nuclei with the same spin momentum, i.e., a nuclear spin can be flopped from "down" to "up" but not vice versa. Because of the long relaxation time of nuclear spins at low temperatures, the flopped nuclear spins are steadily accumulated during many cycles (with a period of ~ 100 ns) of the (1,1) triplet occupation followed by scattering from the (1,1) triplet to the (1,1) singlet, and this eventually leads to dynamical polarization of nuclei. Assuming the electron g-factor of $-0.44$ for our quantum dot, we calculate the magnetic field of $B_{DC} = 0.4$-2 T for making degenerate the (1,1) triplet ($S_Z = +1$) state and the (1,1) singlet state. This calculation agrees with the experimental $B_{DC}$ field where a step and oscillations are observed. A sample with wider center barrier has a smaller tunnel coupling. Therefore, the separation between the (1,1) triplet and (1,1) singlet states at $B_{DC} = 0$ T is smaller and the degeneracy of the two states should occur at a lower $B_{DC}$ field. This quantitatively agrees with our observation that step and oscillations are observed at lower $B_{DC}$ fields for the 7.5-nm-center-barrier sample.

Although any detailed mechanism responsible for the observed current oscillations and their NMR response is not yet understood yet, the oscillations are consistently reproduced if we phenomenologically assume that the larger nuclear polarization leads to the larger period and amplitude of the current oscillations. The nuclear polarization grows with increasing magnetic field (Fig. 2(b)), gradually grows (decays) by turning on (off) the spin blockade (Fig. 2(c)), and resonantly decays under the NMR condition (Fig. 3(a)). These results suggest the presence of complicated back action from the polarized nuclei to the electron spin far beyond the conventional Overhauser effect. It has recently been predicted for our double dot system that the coupled electron-nuclear spin system exhibits instability near the crossover of the (1,1) triplet and (1,1) singlet states [12]. Note in our knowledge oscillations of nuclear polarization with a period of ~100 sec was previously observed in the optical study of bulk *n*-AlGaAs, which referred to nonlinear dynamics of coupled electron-nuclear spin systems [13,14]. Absence of the NMR response from the As nuclei, despite of their largest population, is another open question. The As resonance frequency (6.2 MHz for 0.85 T) can be a singular point for impedance matching of our rf coil. However we rule out this possibility, because for the 7.5-nm-center-barrier sample both of the $^{69}$Ga and $^{71}$Ga signals are observed in the similar frequency range (5.6-7.0 MHz).

In conclusion we have studied magnetic field effects on a small leakage current of order of 1 pA in the spin-blocked vertical double quantum dot system. In the presence of a DC in-plane magnetic field of 0.7-0.87 T we have observed oscillations of the leakage current with a period of as long as 200 sec. Application of the NMR rf field significantly diminishes the oscillatory behavior, indicating the presence of the hyperfine flip-flop scattering and polarized nuclear spin state in the quantum dots.

We thank D. G. Austing, M. Eto, T. Fujisawa, K. Hashimoto, X. Hu, T. Inoshita, M. Stopa, and Y. Tokura for valuable discussions. We acknowledge financial support from a Grant-in-Aid for Scientific Research A (No. 40302799) from the Japan Society for the Promotion of Science and from CREST-JST. K.O. thank the cryogenic center, university of Tokyo for the use of low temperature facilities.

[1] *Mesoscopic Electron Transport*, edited by L.L. Sohn, G. Schön and L. P. Kouwenhoven (Kluwer, Series E345, 1997).
[2] S. Tarucha *et al*, Phys. Rev. Lett. **77**, 3613 (1996). L. P. Kouwenhoven *et al.*, Science **278**, 1788

(1997).

[3] D. Goldhaber-Gordon *et al.*, Nature (London) **391**, 156 (1998).

[4] T. Fujisawa *et al.*, Nature (London) **419**, 278 (2002).

[5] A. V. Khaetskii and Y. V. Nazarov, Phys. Rev. B **61**, 12639 (2000).

[6] S. I. Erlingsson, Y. V. Nazarov, and V. I. Fal'ko, Phys. Rev. B **64**, 195306 (2001).

[7] *Semiconductor Spintronics and Quantum Computation*, edited by D. D. Awschalom, N. Samarth and D. Loss (Springer-Verlag, Gerlin, 2002).

[8] M. Dobers *et al.*, Phys. Rev. Lett. **61**, 1650 (1988); A. Berg *et al.*, Phys. Rev. Lett. **64**, 2653 (1990); K. R.Wald *et al.*, Phys. Rev. Lett. **73**, 1011 (1994); D. C. Dixon, *et. al.*, Phys. Rev. B **56**, 4743 (1997); S. Kronmüller *et al.*, Phys. Rev. Lett. **81**, 2526 (1998); S. Kronmüller *et al.*, *ibid.* **82**, 4070 (1999); K. Hashimoto *et al.*, Physica B **298**, 191 (2001); K. Hashimoto *et al.*, Phys. Rev. Lett. **88**, 176601 (2002); J. H. Smet *et al.*, Nature (London) **415**, 281 (2002); T. Machida, T. Yamazaki, and S. Komiyama, Appl. Phys. Lett. **80**, 4178 (2002); T. Machida *et al.*, *ibid.* **82**, 409 (2003).

[9] K. Ono *et al.*, Science **297**, 1313 (2002).

[10] D. G. Austing *et al.*, Physica B **249-251**, 206 (1998).

[11] Y. Tokura, unpublished. The (1,1) triplet state is lie below the (1,1) singlet. Note that the *ground state* is the (0,2) singlet state.

[12] T. Inoshita, K. Ono, and S. Tarucha, J. Phys. Soc. Jpn. **72**, 183 (2003).

[13] V. G. Fleisher *et al.*, JETP Lett. **21**, 255 (1975).

[14] *Optical Orientation*, edited by F. Meier and B. P. Zakharchenya (North-Holland, Amsterdam, 1984).

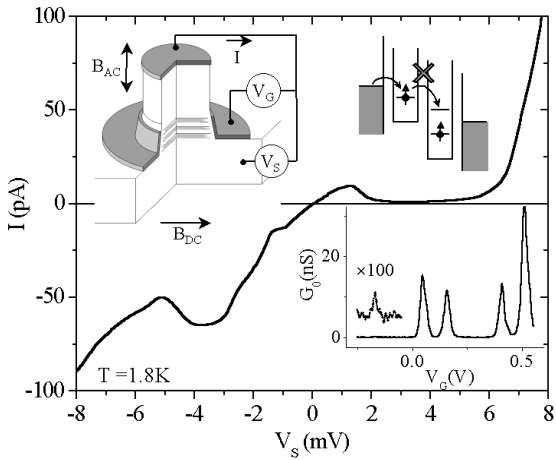

**Fig. 1** Current (*I*)-voltage (*V*$_S$) characteristic measured at *T* = 1.8K. The spin blockade region is present for 2 mV < *V*$_S$ < 6 mV, where the system holds a spin triplet state as illustrated in the upper right

inset. Lower right inset: Gate voltage (*V*$_G$) dependence of the linear conductance. The main *I*-*V*$_S$ curve is taken near the first large peak at *V*$_G$ = 0.05 V, where the three charge states (*N*$_1$,*N*$_2$) = (0,1), (1,1) and (0,2) are degenerated. A small first Coulomb peak, where the (0,1) state is aligned to the Fermi level of the reservoirs, is seen. Left inset: Schematic of the vertical double dot devices. Directions of DC and AC magnetic fields are indicated.

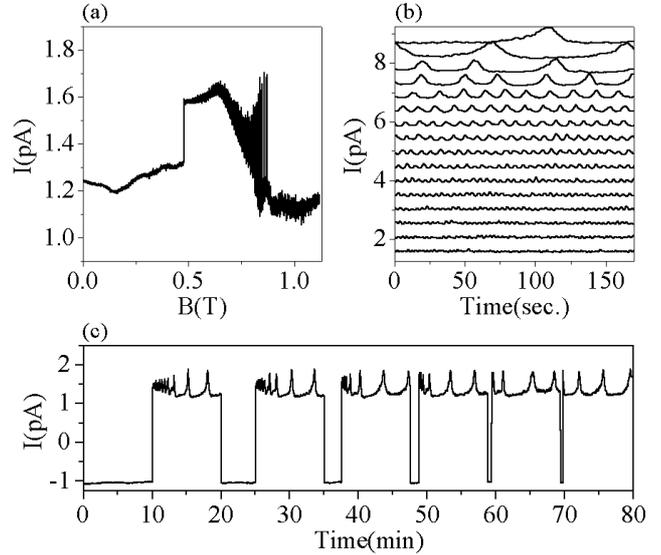

**Fig.2** (a) Magnetic field (*B*$_{DC}$) dependence of the leakage current at *V*$_S$ = 3.0 mV or in the middle of the spin blockade region, as a function of in-plane magnetic field. Detailed positions of the step and largest fluctuations depend on the *B*$_{DC}$ sweep rate and values of *V*$_S$ and *V*$_G$. (b) Leakage current evolving with time measured for fixed magnetic fields of *B*$_{DC}$ = 0.70 T to 0.85 T with 0.01 T step for the curves from bottom to top. Each curve is vertically offset by 0.5 pA for clarity. (c) Transient behavior of the oscillatory current at *B*$_{DC}$ = 0.87 T and *V*$_S$ = 3.0 mV after dwelling at *V*$_S$ = −1.0 mV (*I* = −1 pA) for 600, 300, 150, 75, 36 and 18 sec, respectively.

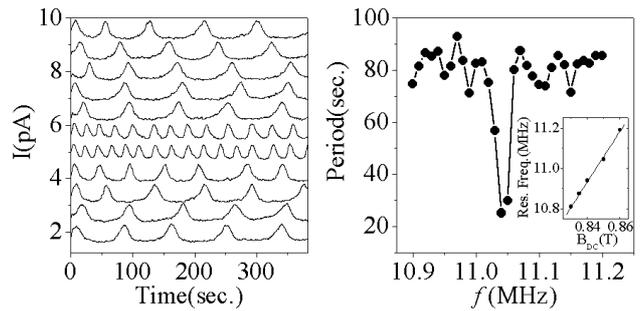

**Fig.3** (a) Current oscillations under AC magnetic field in the frequency range of *f* = 11.00 MHz to 11.10 MHz with 0.01 MHz step for the curves from bottom to top. Each curve is vertically offset by 0.75 pA for clarity. Amplitude of the AC voltage applied to the coil is adjusted so that the effects of neither the heating nor "pumping current" due to the stray electrical coupling to the coil are reasonably small. (b) Time period of the current oscillations measured for various frequencies of the AC magnetic field *f*. Inset: DC magnetic field dependence of the resonance frequency observed in the oscillation period.